\PassOptionsToPackage{cmyk,table}{xcolor}

\makeatletter
\let\blx@rerun@biber\relax
\makeatother

\documentclass[conference,flushend]{iaria}
\usepackage[babel=true,english=american]{csquotes}
\usepackage[british]{babel}
\usepackage[shortcuts]{extdash}
\usepackage[alwaysadjust]{paralist}
\usepackage[babel=true, expansion=alltext, protrusion=alltext-nott, nopatch=eqnum, final]{microtype}
\usepackage{booktabs,array}
\newcolumntype{C}[1]{>{\centering\arraybackslash}p{#1}}
\usepackage{relsize}
\usepackage{pgfplots}
\usepackage{subcaption}
\pgfplotsset{compat=1.18}
\usepackage{amsmath}
\usepackage{balance}
\usepackage{placeins}

\definecolor{haw-blau}{RGB}{0,48,93}
\definecolor{haw-orange}{RGB}{214,123,25}

\title{CVEs With a CVSS Score Greater Than or Equal to 9}

\author{%
    \large Lena Sinterhauf\,$^{1,2}$, Andreas Aßmuth\,$^2$\,\orcidlink{0009-0002-2081-2455}, and Roland Kaltefleiter\,$^1$\\[0.3ex]\normalsize\normalfont
    $^1$\,NetUse AG, Kiel, Germany\\
    e-mail: {\tt \{lsi\,|\,rk\}@netuse.de}\\
    $^2$\,Kiel University of Applied Sciences, Kiel, Germany\\
    e-mail: {\tt andreas.assmuth@haw-kiel.de}
}

\DeclareQuoteStyle{iaria-double}
  {\textquotedblleft}
  {\textquotedblright}
  [0.05em]
  {\textquotedblleft}
  {\textquotedblright}

\setquotestyle{iaria-double}

\usepackage{ifpdf}
\ifpdf
\pdfoutput=1 
\pdftrue
\fi

\makeatletter
\def\ps@IEEEtitlepagestyle{
	\def\@oddfoot{\mycopyrightnotice}
	\def\@evenfoot{}
}
\def\mycopyrightnotice{
	{\footnotesize
		\begin{minipage}{0.8\textwidth}
			\centering
			Please cite as: Lena Sinterhauf, Andreas Aßmuth, and Roland Kaltefleiter, ``CVEs With a CVSS Score Greater Than or Equal to 9,'' in \emph{Proc of the First International Conference on Cross-Domain Security in Distributed, Intelligent and Critical Systems (CROSS-SEC 2026), Lisbon, Portugal, pp.~17--23, April 2026.}
		\end{minipage}
	}
}
\makeatother

\begin{document}

\maketitle

\begin{abstract}
Critical vulnerabilities with Common Vulnerability Scoring System scores of 9.0 or higher pose severe risks to organisations' information systems. Timely detection and remediation are essential to minimise economic and reputational damage from cyberattacks. This paper provides a thorough analysis of the identification and resolution timelines of such critical vulnerabilities. A mixed-methods approach is employed, integrating quantitative data from global vulnerability databases analysing 245{,}456 Common Vulnerabilities and Exposures records spanning from 2009 to 2024, of which 12.8\,\% were critical, with qualitative case studies of notable incidents. This methodical combination of quantitative and qualitative data sources enables the identification of patterns and delay factors in vulnerability management. The findings indicate significant delays in public disclosure and patch deployment, influenced by industry-specific factors, resource availability and organisational processes. The paper concludes with a series of actionable recommendations to improve the efficiency of vulnerability responses. Despite faster disclosure, the remediation gap for critical vulnerabilities remains a systemic risk, driven by organisational inertia and system complexity.
\end{abstract}


\section{Introduction}
The increasing digitisation and interconnectivity of organisations and critical infrastructures has led to a growing threat landscape dominated by cyberattacks. Information security constitutes a central component within contemporary corporate strategies, with the objective of preserving the integrity, availability, and confidentiality of data. The standardised identification and documentation of security vulnerabilities is facilitated by the Common Vulnerabilities and Exposures (CVE) system~\cite{cve_faqs}. The assessment of vulnerabilities is conducted through the utilisation of the Common Vulnerability Scoring System (CVSS), a method that employs a scale ranging from $0$ (lowest) to $10$ (highest) to evaluate the severity of the vulnerabilities identified~\cite{nist_cvss_scoring}. Vulnerabilities that receive a score of $9.0$ or higher are designated as critical, given the substantial risks they pose to the affected systems.\par 

Despite the implementation of established security processes, the timely detection and remediation of critical vulnerabilities remain challenging. The failure to disclose vulnerabilities or the failure to deploy patches in a timely manner can expose organisations to significant risks of exploitation, resulting in financial losses, operational disruptions and reputational damage~\cite{ibm_cost_of_breach, bsi_lage_it-sicherheit_schwachstellen_wachstum}. Notable incidents, such as the Log4Shell vulnerability, have underscored the pressing need for effective and efficient vulnerability management~\cite{nist_cve-2021-44228, bsi_cve-2021-44228, crowdstrike_log4shell}.\par

The present study investigates the speed and efficiency of identifying and remediating critical vulnerabilities, focusing on those with CVSS scores of $9.0$ or above. It synthesises quantitative analyses of vulnerability data from global databases spanning 2009 to 2024 with qualitative case studies of major security incidents. The objective of this study is to identify the factors that contribute to delays in vulnerability response and to provide actionable recommendations for improving the resilience of IT infrastructures against critical security threats.\par

This study provides a long-term analysis of critical vulnerabilities (CVSS $\geq 9.0$) across 245,456~CVE records over a period of 16 years (2009 to 2024). Contrary to previous studies, this research combines large-scale quantitative analysis with qualitative case studies of major incidents (Heartbleed, EternalBlue, and Log4Shell) to identify systemic delays in vulnerability remediation and examines sector-specific patch patterns across more than 20 industries, thus providing practical insights for the prioritisation of vulnerability remediation in the context of limited security resources.

The structure of this paper is as follows: Section~II provides a comprehensive review of the extant literature on vulnerability management and scoring systems. Section~III delineates the research methodology. The fourth section of this text presents the results of the data analyses and case studies. The subsequent section, Section~V, discusses the implications of these findings. Finally, Section~VI concludes with a summary and suggestions for future research.

\section{Related work}\label{chap:related_work}
Research on software vulnerabilities has addressed the subjects of detection, severity assessment, and remediation. CVSS a widely utilised numerical classification system for vulnerability criticality, which serves to guide the prioritisation of remediation efforts~\cite{nist_cvss_scoring, first_cvss_v4_guide}. Service Level Agreements (SLAs) have been proposed as a means of defining remediation timelines. It is recommended that critical vulnerabilities be addressed within days to weeks ~\cite{bsi_patchmanagement_2021, alexiou_patch_management}.\par

Empirical studies have analysed vulnerability lifecycles, management frameworks, open-source processes, and metrics, such as mean time to remediate and disclosure-to-patch delays~\cite{nist_patch_guide_2022, nurse_patch_challenges}. The extant literature consistently highlights challenges in the timely remediation of issues, which are influenced by system complexity, organisational readiness, and resource constraints.

\begin{table*}[!h]
    \centering
    \rowcolors{3}{black!10}{white}
    \begin{tabular}{p{1.2cm}p{4.5cm}p{3.5cm}p{5cm}}
        \toprule
        Year & Title & Author(s) & Topics \\
        \midrule
        2019 & Practical patch management and mitigation & S. Alexiou & Patch SLAs, remediation timelines~\cite{alexiou_patch_management} \\
        2022 & Guide to enterprise patch management planning & M. Souppaya, K. Scarfone & MTTR metrics, enterprise patching~\cite{nist_patch_guide_2022} \\
        2025 & To patch or not to patch & J.R.C. Nurse & Patching motivations, organisational challenges~\cite{nurse_patch_challenges} \\
        2025 & The secret life of CVEs & P. Przymus et al. & CVE lifecycle analysis~\cite{przymus_secret_life_of_cves} \\
        2025 & Out of sight, still at risk & P. Przymus et al. & Transitive vulnerabilities, Maven ecosystem~\cite{przymus_transitive_lifecycle} \\
        \bottomrule
    \end{tabular}
    \caption{Related Work on CVE/CVSS Analysis (2009 to 2025)}
    \label{tab:related_work}
\end{table*}

Research focusing explicitly on critical vulnerabilities reports that, despite improvements in disclosure speed, patch deployment often lags behind, thereby extending exposure windows and exploitation risk~\cite{przymus_secret_life_of_cves, przymus_transitive_lifecycle}. Despite the existence of regulatory and sector-specific guidelines that advocate for faster responses, the efficacy of patching varies across sectors and vendors.

Key studies on vulnerability management and CVE lifecycles have been summarised in Table~\ref{tab:related_work}. Whilst the extant literature provides insights into patching processes and vulnerability lifecycles, none offer a long-term (2009 to 2024) analysis focused on critical vulnerabilities (CVSS $\geq 9.0$) across sectors, combined with qualitative case studies of high-impact incidents. The present study addresses this gap by integrating large-scale quantitative data with detailed case analyses to uncover patterns, delays, and factors unique to the highest severity vulnerabilities.

\section{Methods}
The present study employs a mixed-methods approach in order to comprehensively analyse the identification and remediation processes of critical security vulnerabilities. The methodology integrates quantitative analysis of vulnerability data with qualitative case studies to gain both breadth and depth of understanding.\par

The quantitative component employs data from recognised vulnerability databases, namely the National Vulnerability Database (NVD) and the MITRE CVE database~\cite{nist_cve_json2.0_data, mitre_cve_json_data}. The data were obtained from the official NVD JSON~2.0 feeds (as of 28 May 2025) and MITRE CVE list downloads (as of 20 May 2025). The datasets were processed using Python scripts (e.g., \textit{pandas}, \textit{json}) to filter vulnerabilities with CVSS base scores of $\geq 9.0$, to calculate temporal metrics (e.g., days from \textit{reservation\_date} to \textit{published} and to \textit{lastModifiedDate} as patch proxy), and to aggregate results by year, assignee (assigners), and sector.

The analysis encompasses $245{,}456$ CVE records registered between January~$2009$ and December~$2024$, of which $31{,}430$ (approx. $12.8\,\%$) were classified as critical with CVSS base scores of $9.0$ or higher. Two primary temporal dimensions were examined: the duration from CVE reservation to public disclosure, and the duration from disclosure to patch availability (approximated using database modification timestamps). Statistical analysis was conducted to examine these timelines, identify trends and patterns over time, and reveal discrepancies across industry sectors and software categories.\par 

The qualitative element of the study comprises in-depth case studies of notable security incidents, including the Heartbleed, EternalBlue, and Log4Shell vulnerabilities~\cite{nist_heartbleed_cve, nist_eternalblue_cve, bsi_cve-2021-44228}. The case studies presented offer insights into the challenges encountered by organisations in the field of vulnerability management, including issues, such as delays, organisational factors, and best practices in mitigation.\par

The integration of quantitative and qualitative data facilitates cross-validation and more profound interpretation of results. Quantitative findings reveal statistical trends and potential delay factors, while qualitative analysis contextualises these findings within actual incident scenarios and management practices.\par

The process of data validation entailed several key steps. Firstly, database entries were subjected to rigorous cross-checking. Secondly, the results were meticulously compared with existing literature to ensure the reliability of the findings. Finally, consistency was maintained throughout all analysis phases. This integrated approach facilitates the development of pragmatic recommendations that are designed to enhance the efficiency and effectiveness of critical vulnerability management.

\section{Results}
The analysis encompasses vulnerability data from $2009$ to $2024$, with a particular emphasis on critical vulnerabilities that have a severity score of $9.0$ or higher. The results of the study reveal three primary dimensions of interest: detection and publication timelines, patch availability delays, and organisational or sectoral variation in response times.

\subsection{Detection and Publication Timelines}
\begin{figure}[htbp]
	\centering
	\small
	\begin{tikzpicture}
		\begin{axis}[width=0.95\columnwidth, height=4.5cm, ybar=2pt, bar width=4pt, enlarge x limits=0.05, xlabel={Year of Reservation}, ylabel={Number of CVEs}, xtick=data, xticklabel style={rotate=90, anchor=east, /pgf/number format/.cd, set thousands separator={}}, ymin=0, scaled y ticks=false,	yticklabel style={/pgf/number format/.cd, fixed, set thousands separator={,}}, legend style={at={(0.02,0.98)}, anchor=north west, font=\footnotesize}, legend cell align={left}, ymajorgrids=true, grid style={dashed, gray!30}, legend image code/.code={\draw[#1, fill, yshift=-0.07cm] (0cm,0cm) rectangle (0.2cm,0.2cm);}]
			\addplot[fill=haw-orange, draw=haw-orange, line width=0.3pt]
				coordinates {
					(2009, 3882)
					(2010, 4116)
					(2011, 3944)
					(2012, 4898)
					(2013, 5263)
					(2014, 7993)
					(2015, 8636)
					(2016, 13095)
					(2017, 14504)
					(2018, 14958)
					(2019, 18503)
					(2020, 19698)
					(2021, 18045)
					(2022, 13149)
					(2023, 4809)
					(2024, 4111)
			};
			\addplot[fill=haw-blau, draw=haw-blau, line width=0.4pt] 
				coordinates {
					(2009, 836)
					(2010, 1000)
					(2011, 783)
					(2012, 830)
					(2013, 947)
					(2014, 879)
					(2015, 1343)
					(2016, 1240)
					(2017, 2056)
					(2018, 1841)
					(2019, 2431)
					(2020, 2589)
					(2021, 2277)
					(2022, 2134)
					(2023, 1253)
					(2024, 953)
			};
				
			\legend{All CVEs, Critical CVEs}
		\end{axis}
	\end{tikzpicture}
	\caption{Annual distribution of CVE registrations (2009--2024). Orange bars represent total no. of CVEs, while blue bars show no. of critical vulnerabilities.}
	\label{fig:reserved_cves}
\end{figure}
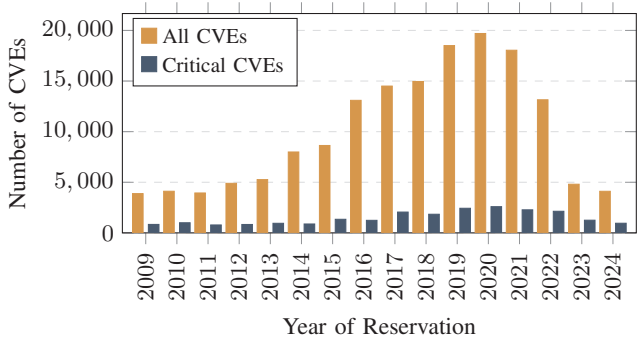

The total number of registered vulnerabilities increased substantially over the study period, particularly after $2016$ when the CVE assignment process was expanded through the introduction of CVE Numbering Authorities (CNAs)~\cite{cve_cna, cve_cna_list}, as illustrated in Figure~\ref{fig:reserved_cves}.

In this context, ``registrations'' denote the initial allocation of a CVE ID by MITRE or designated CNAs upon internal vulnerability reporting, prior to public disclosure~\cite{cve_glossar_record, cve_record_process}. ``Publications'' refer to the subsequent public release of detailed vulnerability information in databases, such as MITRE and the NVD, making it visible to the global security community~\cite{cve_glossar_record, cve_record_process, nist_nvd-process}.

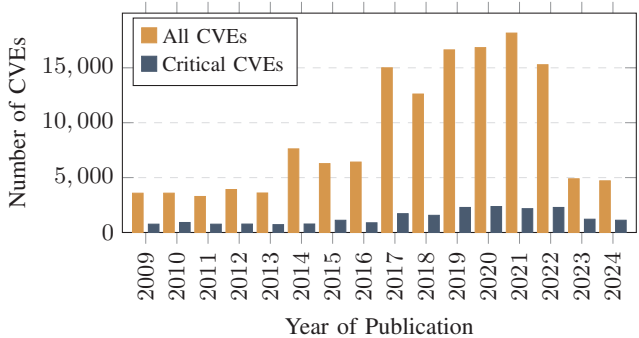
\begin{figure}[htbp]
	\centering
	\small
	\begin{tikzpicture}
		\begin{axis}[width=0.95\columnwidth, height=4.5cm, ybar=2pt, bar width=4pt, enlarge x limits=0.05, xlabel={Year of Publication}, ylabel={Number of CVEs}, xtick=data, xticklabel style={rotate=90, anchor=east, /pgf/number format/.cd, set thousands separator={}}, ymin=0, scaled y ticks=false,	yticklabel style={/pgf/number format/.cd, fixed, set thousands separator={,}},	legend style={at={(0.02,0.98)}, anchor=north west, font=\footnotesize},	legend cell align={left}, ymajorgrids=true,	grid style={dashed, gray!30},legend image code/.code={\draw[#1, fill, yshift=-0.07cm] (0cm,0cm) rectangle (0.2cm,0.2cm);}]
			\addplot[fill=haw-orange, draw=haw-orange, line width=0.3pt] 
				coordinates {
					(2009, 3580)
					(2010, 3583)
					(2011, 3281)
					(2012, 3915)
					(2013, 3598)
					(2014, 7616)
					(2015, 6273)
					(2016, 6416)
					(2017, 15000)
					(2018, 12612)
					(2019, 16633)
					(2020, 16837)
					(2021, 18163)
					(2022, 15280)
					(2023, 4892)
					(2024, 4706)
			};
			\addplot[fill=haw-blau, draw=haw-blau, line width=0.4pt] 
				coordinates {
					(2009, 757)
					(2010, 909)
					(2011, 756)
					(2012, 766)
					(2013, 713)
					(2014, 771)
					(2015, 1112)
					(2016, 882)
					(2017, 1712)
					(2018, 1562)
					(2019, 2279)
					(2020, 2362)
					(2021, 2168)
					(2022, 2279)
					(2023, 1208)
					(2024, 1115)
			};
				
		\legend{All CVEs, Critical CVEs}
		\end{axis}
	\end{tikzpicture}
	\caption{Annual distribution of CVE publications (2009--2024). Orange bars represent total no. of CVEs, while blue bars show no. of critical vulnerabilities.}
	\label{fig:published_cves}
\end{figure}

This phenomenon resulted in a marked increase in the publication of vulnerabilities in $2017$ and again during the period of the pandemic caused by the virus known as SARS-CoV-2 ($2020$–$2021$), when working remotely and accelerated digitisation led to a greater number of exposed systems (cf. Figure~\ref{fig:published_cves}).

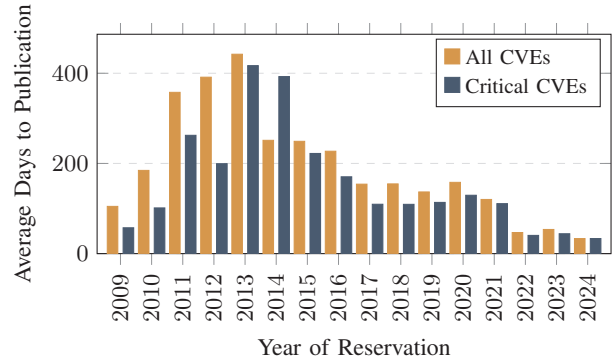
\begin{figure}[htbp]
	\centering
	\small
	\begin{tikzpicture}
		\begin{axis}[width=0.95\columnwidth, height=4.5cm, ybar=2pt, bar width=4pt, enlarge x limits=0.05, xlabel={Year of Reservation}, ylabel={Average Days to Publication}, xtick=data, xticklabel style={rotate=90, anchor=east, /pgf/number format/.cd, set thousands separator={}}, ymin=0, scaled y ticks=false, yticklabel style={/pgf/number format/.cd, fixed, set thousands separator={,}}, legend style={at={(0.98,0.98)}, anchor=north east, font=\footnotesize}, legend cell align={left}, ymajorgrids=true, grid style={dashed, gray!30},legend image code/.code={\draw[#1, fill, yshift=-0.07cm] (0cm,0cm) rectangle (0.2cm,0.2cm);}]
			\addplot[fill=haw-orange, draw=haw-orange, line width=0.3pt] 
			coordinates {
				(2009, 104.8083)
				(2010, 184.6251)
				(2011, 357.7319)
				(2012, 391.329)
				(2013, 442.2771)
				(2014, 251.327)
				(2015, 249.1005)
				(2016, 227.1544)
				(2017, 154.0873)
				(2018, 154.6574)
				(2019, 136.8134)
				(2020, 158.1191)
				(2021, 120.1808)
				(2022, 46.81252)
				(2023, 53.61181)
				(2024, 33.39907)
			};
			\addplot[fill=haw-blau, draw=haw-blau, line width=0.4pt] 
			coordinates {
				(2009, 57.29187)
				(2010, 101.349)
				(2011, 262.376)
				(2012, 199.441)
				(2013, 416.9018)
				(2014, 392.6477)
				(2015, 222.1304)
				(2016, 170.3605)
				(2017, 109.4779)
				(2018, 109.2956)
				(2019, 113.5774)
				(2020, 129.4242)
				(2021, 110.8125)
				(2022, 40.39745)
				(2023, 44.05906)
				(2024, 33.41868)
			};
			
			\legend{All CVEs, Critical CVEs}
		\end{axis}
	\end{tikzpicture}
	\caption{Average time from CVE reservation to public disclosure (2009--2024). Orange bars represent all CVEs, while blue bars show critical vulnerabilities.}
	\label{fig:time_to_publication}
\end{figure}

The proportion of critical vulnerabilities remained relatively stable throughout the observation period at approximately $12.8\,\%$ of all registered CVEs, with a peak of $2{,}589$ cases recorded in $2020$. As illustrated in Figure~\ref{fig:time_to_publication}, the average time from CVE reservation to public disclosure has decreased dramatically, from over $400$~days in $2013$ to approximately $33$~days in $2024$.

Notably, while critical vulnerabilities were published 
significantly faster than the overall average in earlier years (e.g., $57$ vs. $105$~days in $2009$), this gap has virtually disappeared in recent years, indicating that systematic improvements in disclosure processes now benefit all vulnerability severity levels equally. This convergence represents a positive development in the CVE ecosystem. Nevertheless, the CVSS remains essential for severity classification and prioritisation, particularly when organisations face resource constraints or must process large numbers of vulnerabilities simultaneously, making it advisable to address critical issues first.\par 

Our investigations also revealed that the time between registration and publication of a vulnerability varies considerably. While in some cases, assigners published registered CVEs on the same day (duration $0\,\text{days}$), in other cases it took up to several years. Of course, the reasons for these enormous differences in time are not apparent from the data available to us. In cases where the time span is very short, immediate publication is usually due to already known or simultaneously published vulnerabilities that were subsequently assigned a CVE~ID. However, short time spans also indicate that some organisations appear to have particularly efficient processes in place, possibly automated disclosure procedures or internal Standard Operating 
Procedures (SOPs) that prioritise rapid publication. Long durations do not automatically mean that poor work was done in these cases. Reasons for this can also include complex coordination processes, late discovery of the actual impact, or subsequent publication of confidential vulnerabilities~\cite{bsi_cvd_leitlinie, nist_nvd-process}.\par 

Notably, none of the assigners have an average publication time of $0\,\text{days}$ for critical CVEs. This indicates that critical vulnerabilities are always subjected to at least a brief review before they are made public. Furthermore, it can be observed that the longest average delay for critical CVEs (approx. $850\,\text{days}$) is significantly shorter than for all CVEs (over $2{,}300\,\text{days}$). This suggests that critical vulnerabilities are generally processed and published more quickly, even when delays occur. At the same time, the data shows that the variance in duration for critical CVEs is lower, suggesting increased process standardisation or prioritisation.

\subsection{Time to Patch Availability}
Patch deployment analysis demonstrates that turnaround times are both longer and more variable than those observed in the publication phase. The mean time to release a patch for general vulnerabilities was approximately $1{,}732\,\text{days}$ (median: $1{,}335\,\text{days}$), whereas for critical vulnerabilities it was around $2{,}024\,\text{days}$ (median: $1{,}668\,\text{days}$).

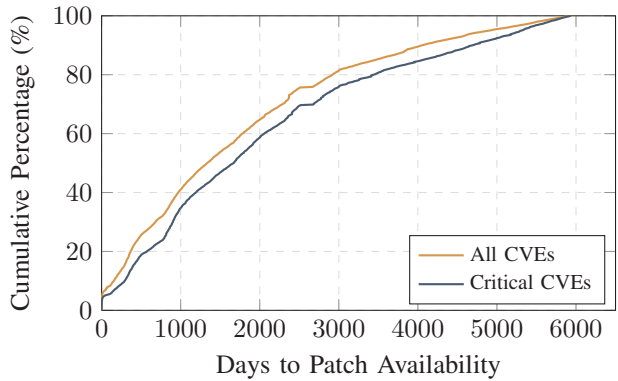
\begin{figure}[htbp]
    \centering
    \begin{tikzpicture}
        \begin{axis}[width=0.95\columnwidth, height=5.5cm, xlabel={Days to Patch Availability}, ylabel={Cumulative Percentage (\%)},
            xmin=0, xmax=6500, ymin=0, ymax=100, xtick={0,1000,2000,3000,4000,5000,6000}, xticklabel style={/pgf/number format/.cd, set thousands separator={}},
            scaled y ticks=false, yticklabel style={/pgf/number format/.cd, fixed}, legend style={at={(0.98,0.02)}, anchor=south east, font=\footnotesize}, legend cell align={left}, xmajorgrids=true, ymajorgrids=true, grid style={dashed, gray!30}, ]

            \addplot[haw-orange, thick, mark=none,] 
                table[col sep=space] {%
                    days cumulative_percent
                    0 0.0004
                    0 0.2000
                    0 0.3997
                    0 0.5993
                    0 0.7989
                    0 0.9985
                    0 1.1982
                    0 1.3978
                    0 1.5974
                    0 1.7971
                    0 1.9967
                    1 2.1963
                    1 2.3959
                    1 2.5956
                    1 2.7952
                    2 2.9948
                    2 3.1945
                    2 3.3941
                    3 3.5937
                    3 3.7933
                    3 3.9930
                    4 4.1926
                    4 4.3922
                    5 4.5919
                    6 4.7915
                    6 4.9911
                    7 5.1907
                    9 5.3904
                    11 5.5900
                    13 5.7896
                    16 5.9893
                    20 6.1889
                    27 6.3885
                    35 6.5881
                    43 6.7878
                    53 6.9874
                    53 7.1870
                    61 7.3867
                    67 7.5863
                    67 7.7859
                    81 7.9855
                    94 8.1852
                    108 8.3848
                    114 8.5844
                    121 8.7841
                    126 8.9837
                    129 9.1833
                    135 9.3829
                    142 9.5826
                    148 9.7822
                    154 9.9818
                    160 10.1815
                    162 10.3811
                    167 10.5807
                    171 10.7803
                    178 10.9800
                    184 11.1796
                    190 11.3792
                    194 11.5789
                    199 11.7785
                    202 11.9781
                    206 12.1777
                    214 12.3774
                    219 12.5770
                    225 12.7766
                    230 12.9763
                    236 13.1759
                    240 13.3755
                    246 13.5751
                    251 13.7748
                    255 13.9744
                    260 14.1740
                    267 14.3737
                    275 14.5733
                    279 14.7729
                    284 14.9725
                    288 15.1722
                    291 15.3718
                    294 15.5714
                    298 15.7711
                    301 15.9707
                    306 16.1703
                    309 16.3699
                    311 16.5696
                    314 16.7692
                    316 16.9688
                    319 17.1685
                    324 17.3681
                    329 17.5677
                    335 17.7673
                    338 17.9670
                    341 18.1666
                    343 18.3662
                    345 18.5659
                    347 18.7655
                    350 18.9651
                    352 19.1647
                    355 19.3644
                    359 19.5640
                    364 19.7636
                    367 19.9633
                    370 20.1629
                    374 20.3625
                    378 20.5621
                    382 20.7618
                    386 20.9614
                    388 21.1610
                    392 21.3607
                    395 21.5603
                    400 21.7599
                    404 21.9595
                    408 22.1592
                    414 22.3588
                    419 22.5584
                    424 22.7581
                    429 22.9577
                    434 23.1573
                    440 23.3569
                    444 23.5566
                    451 23.7562
                    457 23.9558
                    462 24.1554
                    468 24.3551
                    471 24.5547
                    476 24.7543
                    483 24.9540
                    488 25.1536
                    494 25.3532
                    499 25.5528
                    504 25.7525
                    513 25.9521
                    523 26.1517
                    531 26.3514
                    539 26.5510
                    551 26.7506
                    562 26.9502
                    569 27.1499
                    580 27.3495
                    587 27.5491
                    595 27.7488
                    603 27.9484
                    614 28.1480
                    622 28.3476
                    629 28.5473
                    635 28.7469
                    644 28.9465
                    652 29.1462
                    659 29.3458
                    664 29.5454
                    673 29.7450
                    680 29.9447
                    686 30.1443
                    688 30.3439
                    695 30.5436
                    706 30.7432
                    713 30.9428
                    721 31.1424
                    733 31.3421
                    744 31.5417
                    754 31.7413
                    766 31.9410
                    774 32.1406
                    779 32.3402
                    786 32.5398
                    794 32.7395
                    799 32.9391
                    804 33.1387
                    807 33.3384
                    814 33.5380
                    818 33.7376
                    821 33.9372
                    827 34.1369
                    832 34.3365
                    838 34.5361
                    844 34.7358
                    847 34.9354
                    852 35.1350
                    855 35.3346
                    860 35.5343
                    864 35.7339
                    868 35.9335
                    874 36.1332
                    878 36.3328
                    882 36.5324
                    889 36.7320
                    891 36.9317
                    897 37.1313
                    902 37.3309
                    904 37.5306
                    910 37.7302
                    916 37.9298
                    920 38.1294
                    924 38.3291
                    930 38.5287
                    935 38.7283
                    941 38.9280
                    946 39.1276
                    951 39.3272
                    956 39.5268
                    963 39.7265
                    967 39.9261
                    972 40.1257
                    978 40.3254
                    983 40.5250
                    989 40.7246
                    995 40.9242
                    1006 41.1239
                    1011 41.3235
                    1015 41.5231
                    1022 41.7228
                    1029 41.9224
                    1035 42.1220
                    1039 42.3216
                    1044 42.5213
                    1055 42.7209
                    1061 42.9205
                    1067 43.1202
                    1072 43.3198
                    1079 43.5194
                    1086 43.7190
                    1096 43.9187
                    1101 44.1183
                    1110 44.3179
                    1118 44.5176
                    1126 44.7172
                    1132 44.9168
                    1140 45.1164
                    1148 45.3161
                    1157 45.5157
                    1163 45.7153
                    1169 45.9150
                    1176 46.1146
                    1183 46.3142
                    1190 46.5138
                    1196 46.7135
                    1203 46.9131
                    1212 47.1127
                    1219 47.3123
                    1226 47.5120
                    1234 47.7116
                    1245 47.9112
                    1254 48.1109
                    1260 48.3105
                    1269 48.5101
                    1279 48.7097
                    1288 48.9094
                    1295 49.1090
                    1307 49.3086
                    1315 49.5083
                    1323 49.7079
                    1330 49.9075
                    1338 50.1071
                    1350 50.3068
                    1358 50.5064
                    1370 50.7060
                    1379 50.9057
                    1384 51.1053
                    1394 51.3049
                    1401 51.5045
                    1407 51.7042
                    1412 51.9038
                    1421 52.1034
                    1434 52.3031
                    1442 52.5027
                    1454 52.7023
                    1465 52.9019
                    1470 53.1016
                    1479 53.3012
                    1491 53.5008
                    1496 53.7005
                    1506 53.9001
                    1519 54.0997
                    1526 54.2993
                    1533 54.4990
                    1546 54.6986
                    1556 54.8982
                    1568 55.0979
                    1582 55.2975
                    1590 55.4971
                    1603 55.6967
                    1612 55.8964
                    1620 56.0960
                    1627 56.2956
                    1640 56.4953
                    1652 56.6949
                    1666 56.8945
                    1673 57.0941
                    1680 57.2938
                    1681 57.4934
                    1691 57.6930
                    1702 57.8927
                    1707 58.0923
                    1711 58.2919
                    1716 58.4915
                    1728 58.6912
                    1740 58.8908
                    1745 59.0904
                    1749 59.2901
                    1758 59.4897
                    1765 59.6893
                    1772 59.8889
                    1784 60.0886
                    1798 60.2882
                    1801 60.4878
                    1812 60.6875
                    1822 60.8871
                    1833 61.0867
                    1839 61.2863
                    1849 61.4860
                    1862 61.6856
                    1869 61.8852
                    1878 62.0849
                    1883 62.2845
                    1895 62.4841
                    1904 62.6837
                    1916 62.8834
                    1920 63.0830
                    1930 63.2826
                    1937 63.4823
                    1945 63.6819
                    1952 63.8815
                    1963 64.0811
                    1973 64.2808
                    1986 64.4804
                    1997 64.6800
                    2009 64.8797
                    2017 65.0793
                    2031 65.2789
                    2040 65.4785
                    2053 65.6782
                    2060 65.8778
                    2061 66.0774
                    2063 66.2771
                    2074 66.4767
                    2086 66.6763
                    2101 66.8759
                    2114 67.0756
                    2129 67.2752
                    2135 67.4748
                    2142 67.6745
                    2158 67.8741
                    2169 68.0737
                    2179 68.2733
                    2197 68.4730
                    2208 68.6726
                    2224 68.8722
                    2231 69.0718
                    2243 69.2715
                    2254 69.4711
                    2265 69.6707
                    2277 69.8704
                    2294 70.0700
                    2304 70.2696
                    2316 70.4692
                    2323 70.6689
                    2326 70.8685
                    2332 71.0681
                    2346 71.2678
                    2354 71.4674
                    2359 71.6670
                    2365 71.8666
                    2370 72.0663
                    2370 72.2659
                    2370 72.4655
                    2370 72.6652
                    2370 72.8648
                    2371 73.0644
                    2382 73.2640
                    2399 73.4637
                    2408 73.6633
                    2415 73.8629
                    2423 74.0626
                    2433 74.2622
                    2443 74.4618
                    2455 74.6614
                    2468 74.8611
                    2473 75.0607
                    2486 75.2603
                    2499 75.4600
                    2507 75.6596
                    2668 75.8592
                    2683 76.0588
                    2692 76.2585
                    2708 76.4581
                    2718 76.6577
                    2734 76.8574
                    2740 77.0570
                    2755 77.2566
                    2767 77.4562
                    2776 77.6559
                    2789 77.8555
                    2801 78.0551
                    2811 78.2548
                    2819 78.4544
                    2832 78.6540
                    2844 78.8536
                    2858 79.0533
                    2871 79.2529
                    2888 79.4525
                    2899 79.6522
                    2914 79.8518
                    2921 80.0514
                    2931 80.2510
                    2941 80.4507
                    2955 80.6503
                    2965 80.8499
                    2980 81.0496
                    2991 81.2492
                    3004 81.4488
                    3012 81.6484
                    3025 81.8481
                    3052 82.0477
                    3088 82.2473
                    3110 82.4470
                    3132 82.6466
                    3171 82.8462
                    3192 83.0458
                    3213 83.2455
                    3245 83.4451
                    3263 83.6447
                    3286 83.8444
                    3320 84.0440
                    3359 84.2436
                    3381 84.4432
                    3409 84.6429
                    3441 84.8425
                    3465 85.0421
                    3490 85.2418
                    3518 85.4414
                    3542 85.6410
                    3565 85.8406
                    3594 86.0403
                    3622 86.2399
                    3654 86.4395
                    3685 86.6392
                    3717 86.8388
                    3738 87.0384
                    3763 87.2380
                    3786 87.4377
                    3815 87.6373
                    3828 87.8369
                    3831 88.0366
                    3847 88.2362
                    3859 88.4358
                    3868 88.6354
                    3904 88.8351
                    3932 89.0347
                    3959 89.2343
                    3986 89.4340
                    4011 89.6336
                    4033 89.8332
                    4058 90.0328
                    4089 90.2325
                    4111 90.4321
                    4144 90.6317
                    4182 90.8314
                    4208 91.0310
                    4234 91.2306
                    4276 91.4302
                    4311 91.6299
                    4348 91.8295
                    4384 92.0291
                    4420 92.2287
                    4454 92.4284
                    4494 92.6280
                    4533 92.8276
                    4566 93.0273
                    4590 93.2269
                    4609 93.4265
                    4626 93.6261
                    4654 93.8258
                    4689 94.0254
                    4726 94.2250
                    4780 94.4247
                    4817 94.6243
                    4864 94.8239
                    4912 95.0235
                    4943 95.2232
                    4986 95.4228
                    5033 95.6224
                    5087 95.8221
                    5135 96.0217
                    5173 96.2213
                    5207 96.4209
                    5259 96.6206
                    5293 96.8202
                    5341 97.0198
                    5385 97.2195
                    5417 97.4191
                    5452 97.6187
                    5483 97.8183
                    5511 98.0180
                    5559 98.2176
                    5595 98.4172
                    5640 98.6169
                    5676 98.8165
                    5718 99.0161
                    5761 99.2157
                    5800 99.4154
                    5839 99.6150
                    5889 99.8146
                    5937 100.0000
                };
                \addplot[haw-blau, thick, mark=none,] 
                table[col sep=space] {
                    days cumulative_percent
                    0 0.0032
                    0 0.1941
                    0 0.3850
                    0 0.5759
                    0 0.7668
                    0 0.9577
                    0 1.1486
                    1 1.3395
                    1 1.5304
                    1 1.7213
                    1 1.9122
                    1 2.1031
                    2 2.2940
                    2 2.4849
                    2 2.6758
                    3 2.8667
                    3 3.0576
                    4 3.2485
                    5 3.4394
                    6 3.6303
                    7 3.8212
                    9 4.0121
                    12 4.2030
                    15 4.3939
                    19 4.5848
                    29 4.7757
                    38 4.9666
                    55 5.1575
                    81 5.3484
                    104 5.5393
                    117 5.7302
                    125 5.9211
                    132 6.1120
                    136 6.3029
                    145 6.4938
                    151 6.6847
                    160 6.8756
                    164 7.0665
                    174 7.2574
                    188 7.4483
                    192 7.6392
                    202 7.8301
                    208 8.0210
                    217 8.2119
                    224 8.4028
                    235 8.5937
                    244 8.7846
                    252 8.9755
                    260 9.1664
                    269 9.3573
                    280 9.5482
                    286 9.7391
                    289 9.9300
                    295 10.1209
                    300 10.3118
                    304 10.5027
                    309 10.6936
                    314 10.8845
                    316 11.0754
                    321 11.2663
                    325 11.4572
                    329 11.6481
                    335 11.8390
                    337 12.0299
                    341 12.2208
                    343 12.4117
                    345 12.6026
                    350 12.7935
                    352 12.9844
                    357 13.1753
                    359 13.3662
                    364 13.5571
                    370 13.7480
                    377 13.9389
                    380 14.1298
                    384 14.3207
                    388 14.5116
                    392 14.7025
                    393 14.8934
                    399 15.0843
                    402 15.2752
                    407 15.4661
                    415 15.6570
                    421 15.8479
                    426 16.0388
                    428 16.2297
                    434 16.4206
                    441 16.6115
                    448 16.8024
                    451 16.9933
                    455 17.1842
                    462 17.3751
                    465 17.5660
                    470 17.7569
                    474 17.9478
                    478 18.1387
                    485 18.3296
                    493 18.5205
                    498 18.7114
                    503 18.9023
                    515 19.0932
                    524 19.2841
                    538 19.4750
                    552 19.6659
                    569 19.8568
                    573 20.0477
                    584 20.2386
                    594 20.4295
                    604 20.6204
                    616 20.8113
                    626 21.0022
                    633 21.1931
                    643 21.3840
                    652 21.5749
                    659 21.7658
                    664 21.9567
                    676 22.1476
                    683 22.3385
                    693 22.5294
                    706 22.7203
                    716 22.9112
                    729 23.1021
                    737 23.2930
                    748 23.4839
                    761 23.6748
                    773 23.8657
                    779 24.0566
                    783 24.2475
                    791 24.4384
                    793 24.6293
                    796 24.8202
                    799 25.0111
                    805 25.2020
                    811 25.3929
                    815 25.5838
                    817 25.7747
                    818 25.9656
                    819 26.1565
                    824 26.3474
                    828 26.5383
                    833 26.7292
                    838 26.9201
                    842 27.1110
                    846 27.3019
                    849 27.4928
                    854 27.6837
                    860 27.8746
                    864 28.0655
                    864 28.2564
                    868 28.4473
                    874 28.6382
                    880 28.8291
                    884 29.0200
                    888 29.2109
                    889 29.4018
                    893 29.5927
                    897 29.7836
                    902 29.9745
                    903 30.1654
                    904 30.3563
                    909 30.5472
                    913 30.7381
                    917 30.9290
                    921 31.1199
                    923 31.3108
                    925 31.5017
                    929 31.6927
                    932 31.8836
                    937 32.0745
                    942 32.2654
                    944 32.4563
                    951 32.6472
                    954 32.8381
                    960 33.0290
                    966 33.2199
                    969 33.4108
                    973 33.6017
                    978 33.7926
                    980 33.9835
                    983 34.1744
                    987 34.3653
                    994 34.5562
                    1001 34.7471
                    1006 34.9380
                    1013 35.1289
                    1016 35.3198
                    1021 35.5107
                    1027 35.7016
                    1034 35.8925
                    1042 36.0834
                    1055 36.2743
                    1061 36.4652
                    1063 36.6561
                    1067 36.8470
                    1073 37.0379
                    1079 37.2288
                    1084 37.4197
                    1093 37.6106
                    1104 37.8015
                    1110 37.9924
                    1114 38.1833
                    1119 38.3742
                    1127 38.5651
                    1136 38.7560
                    1146 38.9469
                    1153 39.1378
                    1163 39.3287
                    1169 39.5196
                    1178 39.7105
                    1188 39.9014
                    1196 40.0923
                    1201 40.2832
                    1206 40.4741
                    1217 40.6650
                    1225 40.8559
                    1232 41.0468
                    1245 41.2377
                    1254 41.4286
                    1262 41.6195
                    1274 41.8104
                    1282 42.0013
                    1294 42.1922
                    1302 42.3831
                    1308 42.5740
                    1317 42.7649
                    1328 42.9558
                    1337 43.1467
                    1350 43.3376
                    1360 43.5285
                    1370 43.7194
                    1379 43.9103
                    1384 44.1012
                    1392 44.2921
                    1400 44.4830
                    1407 44.6739
                    1416 44.8648
                    1421 45.0557
                    1428 45.2466
                    1435 45.4375
                    1441 45.6284
                    1450 45.8193
                    1463 46.0102
                    1472 46.2011
                    1483 46.3920
                    1491 46.5829
                    1496 46.7738
                    1510 46.9647
                    1518 47.1556
                    1527 47.3465
                    1539 47.5374
                    1542 47.7283
                    1556 47.9192
                    1573 48.1101
                    1577 48.3010
                    1588 48.4919
                    1597 48.6828
                    1608 48.8737
                    1615 49.0646
                    1624 49.2555
                    1632 49.4464
                    1647 49.6373
                    1658 49.8282
                    1669 50.0191
                    1678 50.2100
                    1680 50.4009
                    1687 50.5918
                    1693 50.7827
                    1701 50.9736
                    1702 51.1645
                    1708 51.3554
                    1713 51.5463
                    1721 51.7372
                    1731 51.9281
                    1738 52.1190
                    1744 52.3099
                    1751 52.5008
                    1759 52.6917
                    1764 52.8826
                    1776 53.0735
                    1782 53.2644
                    1797 53.4553
                    1799 53.6462
                    1805 53.8371
                    1813 54.0280
                    1825 54.2189
                    1833 54.4098
                    1841 54.6007
                    1848 54.7916
                    1855 54.9825
                    1862 55.1734
                    1869 55.3643
                    1877 55.5552
                    1888 55.7461
                    1896 55.9370
                    1904 56.1279
                    1913 56.3188
                    1916 56.5097
                    1919 56.7006
                    1924 56.8915
                    1932 57.0824
                    1941 57.2733
                    1947 57.4642
                    1955 57.6551
                    1963 57.8460
                    1969 58.0369
                    1981 58.2278
                    1989 58.4187
                    1997 58.6096
                    2007 58.8005
                    2008 58.9914
                    2014 59.1823
                    2023 59.3732
                    2035 59.5641
                    2043 59.7550
                    2053 59.9459
                    2063 60.1368
                    2071 60.3277
                    2084 60.5186
                    2094 60.7095
                    2108 60.9004
                    2116 61.0913
                    2135 61.2822
                    2143 61.4731
                    2157 61.6640
                    2162 61.8549
                    2177 62.0458
                    2187 62.2367
                    2200 62.4276
                    2214 62.6185
                    2225 62.8094
                    2234 63.0003
                    2245 63.1912
                    2252 63.3821
                    2266 63.5730
                    2276 63.7639
                    2287 63.9548
                    2301 64.1457
                    2312 64.3366
                    2318 64.5275
                    2325 64.7184
                    2327 64.9093
                    2337 65.1002
                    2347 65.2911
                    2354 65.4820
                    2354 65.6729
                    2359 65.8638
                    2367 66.0547
                    2380 66.2456
                    2395 66.4365
                    2408 66.6274
                    2408 66.8183
                    2408 67.0092
                    2409 67.2001
                    2415 67.3910
                    2422 67.5819
                    2428 67.7728
                    2437 67.9637
                    2446 68.1546
                    2457 68.3455
                    2466 68.5364
                    2470 68.7273
                    2477 68.9182
                    2483 69.1091
                    2492 69.3000
                    2504 69.4909
                    2518 69.6818
                    2677 69.8727
                    2684 70.0636
                    2686 70.2545
                    2696 70.4454
                    2710 70.6363
                    2718 70.8272
                    2728 71.0181
                    2736 71.2090
                    2748 71.3999
                    2762 71.5908
                    2773 71.7817
                    2774 71.9726
                    2776 72.1635
                    2791 72.3544
                    2801 72.5453
                    2801 72.7362
                    2812 72.9271
                    2823 73.1180
                    2832 73.3089
                    2843 73.4998
                    2854 73.6907
                    2867 73.8816
                    2886 74.0725
                    2889 74.2634
                    2906 74.4543
                    2920 74.6452
                    2929 74.8361
                    2938 75.0270
                    2949 75.2179
                    2963 75.4088
                    2984 75.5997
                    2993 75.7906
                    3004 75.9815
                    3008 76.1724
                    3021 76.3633
                    3044 76.5542
                    3090 76.7451
                    3102 76.9360
                    3112 77.1269
                    3130 77.3178
                    3170 77.5087
                    3187 77.6997
                    3213 77.8906
                    3247 78.0815
                    3258 78.2724
                    3258 78.4633
                    3281 78.6542
                    3315 78.8451
                    3355 79.0360
                    3381 79.2269
                    3411 79.4178
                    3411 79.6087
                    3412 79.7996
                    3439 79.9905
                    3462 80.1814
                    3475 80.3723
                    3490 80.5632
                    3503 80.7541
                    3528 80.9450
                    3539 81.1359
                    3559 81.3268
                    3568 81.5177
                    3594 81.7086
                    3622 81.8995
                    3648 82.0904
                    3670 82.2813
                    3704 82.4722
                    3718 82.6631
                    3746 82.8540
                    3775 83.0449
                    3808 83.2358
                    3844 83.4267
                    3867 83.6176
                    3895 83.8085
                    3930 83.9994
                    3958 84.1903
                    3958 84.3812
                    4000 84.5721
                    4031 84.7630
                    4054 84.9539
                    4081 85.1448
                    4106 85.3357
                    4139 85.5266
                    4167 85.7175
                    4199 85.9084
                    4229 86.0993
                    4238 86.2902
                    4276 86.4811
                    4299 86.6720
                    4320 86.8629
                    4347 87.0538
                    4358 87.2447
                    4383 87.4356
                    4428 87.6265
                    4440 87.8174
                    4459 88.0083
                    4473 88.1992
                    4510 88.3901
                    4535 88.5810
                    4565 88.7719
                    4584 88.9628
                    4599 89.1537
                    4617 89.3446
                    4642 89.5355
                    4647 89.7264
                    4681 89.9173
                    4707 90.1082
                    4724 90.2991
                    4752 90.4900
                    4781 90.6809
                    4782 90.8718
                    4804 91.0627
                    4839 91.2536
                    4867 91.4445
                    4899 91.6354
                    4922 91.8263
                    4951 92.0172
                    4984 92.2081
                    4993 92.3990
                    5032 92.5899
                    5047 92.7808
                    5068 92.9717
                    5091 93.1626
                    5144 93.3535
                    5173 93.5444
                    5173 93.7353
                    5194 93.9262
                    5223 94.1171
                    5235 94.3080
                    5268 94.4989
                    5285 94.6898
                    5293 94.8807
                    5321 95.0716
                    5340 95.2625
                    5347 95.4534
                    5363 95.6443
                    5392 95.8352
                    5413 96.0261
                    5419 96.2170
                    5446 96.4079
                    5475 96.5988
                    5491 96.7897
                    5527 96.9806
                    5554 97.1715
                    5590 97.3624
                    5601 97.5533
                    5633 97.7442
                    5650 97.9351
                    5675 98.1260
                    5695 98.3169
                    5732 98.5078
                    5754 98.6987
                    5781 98.8896
                    5802 99.0805
                    5823 99.2714
                    5852 99.4623
                    5878 99.6532
                    5908 99.8441
                    5935 100.0000
                };
            \legend{All CVEs, Critical CVEs}
        \end{axis}
    \end{tikzpicture}
    \caption{Cumulative distribution of time to patch availability (2009--2024). The curves show the percentage of CVEs patched within a given timeframe. Mean values: approximately $1{,}732\,\text{days}$ (all CVEs) and $2{,}024\,\text{days}$ (critical CVEs).}
    \label{fig:cdf_patch_duration}
\end{figure}

Figure~\ref{fig:cdf_patch_duration} illustrates the cumulative distribution of time to patch availability across the entire observation period. Notably, $50\,\%$ of all CVEs received patches within $1{,}335\,\text{days}$ of disclosure, while $90\,\%$ were patched within $4{,}054\,\text{days}$. For critical vulnerabilities, the corresponding values were $1{,}668$ and $4{,}689\,\text{days}$. The distribution demonstrates that despite prioritisation efforts, critical vulnerabilities exhibit longer remediation times on average than the general population, likely reflecting the increased complexity and coordination requirements associated with high-severity issues. The highly skewed distribution, with substantial differences between median and mean values, indicates that while the majority of vulnerabilities are addressed within reasonable timeframes, a significant tail of delayed patches persists across both categories.\par 

\begin{figure}[htbp]
	\centering
	\small
	\begin{tikzpicture}
		\begin{axis}[
			width=0.95\columnwidth, height=6cm,	xlabel={Year}, ylabel={Days to Patch Availability}, xmin=2008.5, xmax=2024.5, ymin=0, xtick={2009,2010,...,2024}, xticklabel style={rotate=90, anchor=east, /pgf/number format/.cd, set thousands separator={}},	scaled y ticks=false, yticklabel style={/pgf/number format/.cd, fixed, set thousands separator={,}}, legend style={at={(0.98,0.98)}, anchor=north east, font=\footnotesize}, legend cell align={left},	ymajorgrids=true, grid style={dashed, gray!30},]
			\addplot+[haw-orange, mark=*, mark options={fill=haw-orange, draw=haw-orange}, error bars/.cd, y dir=both, y explicit,]
			table[x=Year, y=Mean, y error minus expr=\thisrow{Mean}-\thisrow{Min}, y error plus expr=\thisrow{Max}-\thisrow{Mean}] {
				Year	Mean	Min		Max
				2009	5746	5059	5937
				2010	5399	4716	5573
				2011	5031	4366	5211
				2012	4650	4029	4847
				2013	4288	3602	4482
				2014	3915	3236	4117
				2015	3567	2882	3754
				2016	3202	2502	3389
				2017	2765	2139	3058
				2018	2317	1774	2679
				2019	1928	1407	2291
				2020	1593	1040	1923
				2021	1206	675		1579
				2022	903		311		1237
				2023	524		0		860
				2024	185		0		504
				2025	25		0		137
			};
			\addplot+[haw-blau, mark=*, mark options={fill=haw-blau, draw=haw-blau}, error bars/.cd, y dir=both, y explicit,]
			table[x=Year, y=Mean, y error minus expr=\thisrow{Mean}-\thisrow{Min}, y error plus expr=\thisrow{Max}-\thisrow{Mean}] {
				Year	Mean	Min     Max
				2009	5749	5577	5935
				2010	5382	5215	5573
				2011	5037	4849	5211
				2012	4667	4488	4846
				2013	4315	4118	4481
				2014	3941	3756	4114
				2015	3545	3389	3754
				2016	3201	3023	3387
				2017	2834	2666	3025
				2018	2349	2151	2667
				2019	1957	1786	2249
				2020	1614	1420	1909
				2021	1235	1055	1550
				2022	923		708		1236
				2023	545		325		836
				2024	176		0		499
				2025	16		0		133
			};
			\legend{All CVEs, Critical CVEs}
		\end{axis}
	\end{tikzpicture}
	\caption{Patch availability for all CVEs and critical CVEs (2009--2024). The average, minimum and maximum durations are specified for each year. In the trend lines, the orange line indicates all CVEs, while the blue line indicates critical CVEs.}
	\label{fig:patch_availability}
\end{figure}
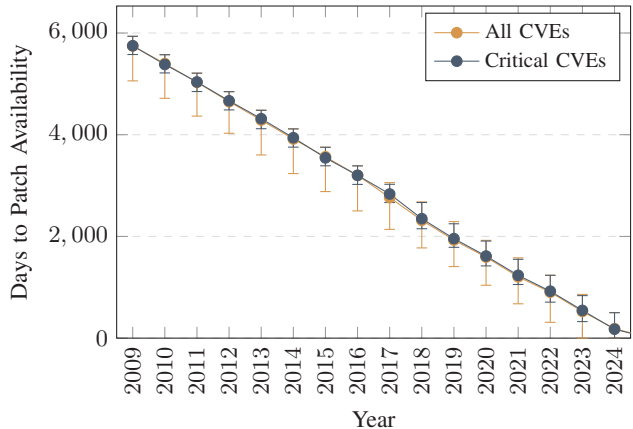

Despite the improvements that have been made, no consistent or significant difference has been demonstrated between critical and non-critical issues with regard to patch completion times, as can be seen in Figure~\ref{fig:patch_availability}. The apparent improvement in recent years should be interpreted with caution due to right-censoring: vulnerabilities from 2022 onwards have had less opportunity to exhibit extended patch delays, and currently unpatched vulnerabilities are not 
represented in these measurements.

Many organisations implement release cycles that are similar across all severity levels. The phenomenon of extended exposure periods can be attributed to various factors, including the increasing complexity of systems, the presence of legacy dependencies, constraints in resources, and the incomplete automation of processes. However, between $2020$ and $2024$, an observable acceleration occurred, suggesting stronger regulatory and procedural pressure on vendors.

\subsection{Organizational and Sectoral Variations} \label{subsec:organizational_and_sectorial_veriations}
\begin{figure*}[!h]
	\centering
	\small
	\begin{tikzpicture}
		\begin{axis}[width=0.95\textwidth, height=6cm, ybar=2pt, bar width=5pt, enlarge x limits=0.03, xlabel={Sector}, ylabel={Median Days to Patch}, ymin=1, xticklabel style={rotate=45, anchor=east, font=\small}, scaled y ticks=false, yticklabel style={/pgf/number format/.cd, fixed}, legend style={at={(0.02,0.98)}, anchor=north west, font=\footnotesize}, legend cell align={left}, ymajorgrids=true, grid style={dashed, gray!30}, symbolic x coords={
            Healthcare,
            Finance \& Insurance,
            Web \& Content Management,
            Other,
            Platforms \& DevOps,
            Industrial \& IoT,
            Consumer Electronics,
            Education \& Non-Profit,
            Hardware,
            Security Vendors,
            Open Source,
            Telecommunications \& Networking,
            Cloud \& Hosting,
            Consulting \& Research,
            Commercial Software
		}, xtick=data, legend image code/.code={\draw[#1, fill, yshift=-0.07cm] (0cm,0cm) rectangle (0.2cm,0.2cm);},]
		      \addplot[fill=haw-orange, draw=haw-orange, line width=0.3pt] 
			coordinates {
				(Healthcare,10)
                (Finance \& Insurance,248)
                (Web \& Content Management,343)
                (Other,407)
                (Platforms \& DevOps,637)
                (Industrial \& IoT,812)
                (Consumer Electronics,944)
                (Education \& Non-Profit,1003)
                (Hardware,1020)
                (Security Vendors,1064)
                (Open Source,1184)
                (Telecommunications \& Networking,1505)
                (Cloud \& Hosting,1744)
                (Consulting \& Research,1917)
                (Commercial Software,1969)
			};
			
			\addplot[fill=haw-blau, draw=haw-blau, line width=0.4pt] 
			coordinates {
				(Healthcare,3)
                (Finance \& Insurance,796)
                (Web \& Content Management,342)
                (Other,889)
                (Platforms \& DevOps,889)
                (Industrial \& IoT,833)
                (Consumer Electronics,1161)
                (Education \& Non-Profit,1924)
                (Hardware,2014)
                (Security Vendors,1444)
                (Open Source,1134)
                (Telecommunications \& Networking,1192)
                (Cloud \& Hosting,3169)
                (Consulting \& Research,1542)
                (Commercial Software,3648)
			};
			
			\legend{All CVEs, Critical CVEs}
		\end{axis}
	\end{tikzpicture}
	\caption{Median time to patch availability by sector (2009 to 2024). Sectors are sorted by overall patch performance (all CVEs). Orange bars represent all CVEs, while blue bars show critical vulnerabilities, revealing significant performance disparities across industries.}
	\label{fig:patch_by_sector}
\end{figure*}
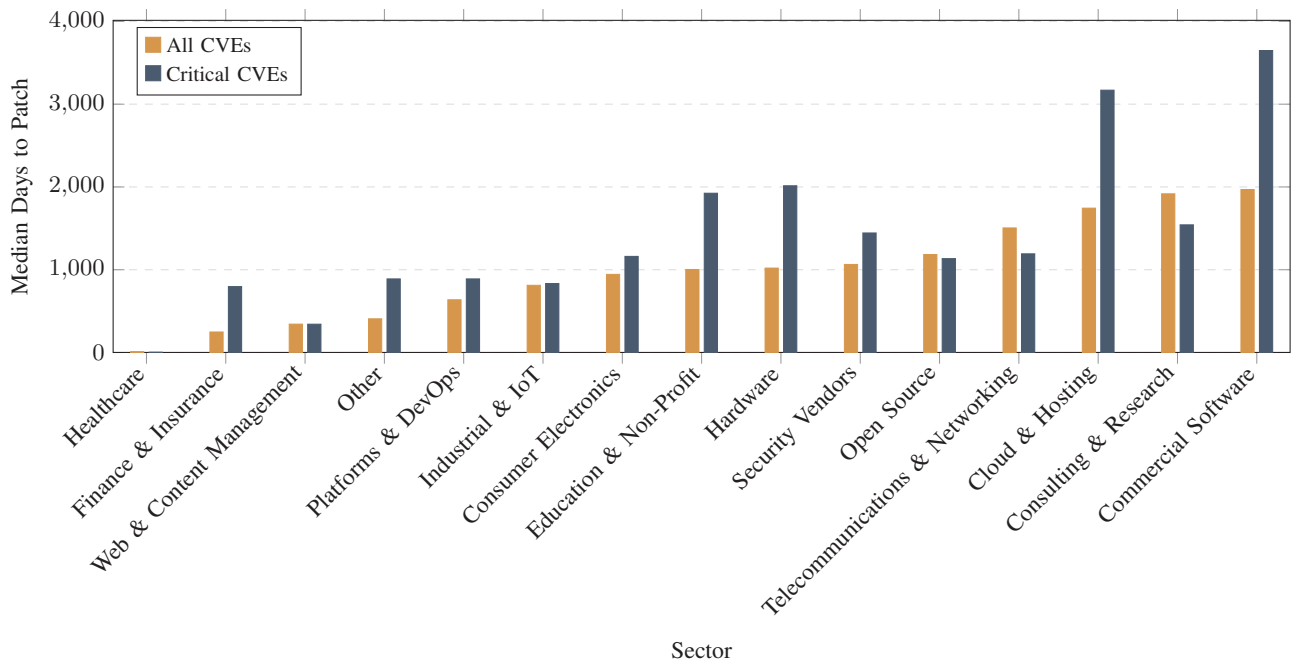

There are notable disparities between different assignees (assigners).  It is evident that certain entities, including specialised security platforms and prominent open-source providers, attained a median patch time of less than five days. This is indicative of sophisticated automation and continuous integration processes. In contrast, slower assigners~-- which were often commercial software vendors~–- exhibited median delays of between $2{,}000$ and $4{,}000\,\text{days}$ (cf. Figure~\ref{fig:patch_by_sector}).

Sectors were assigned by mapping CVE assigners to primary industry classifications using a reproducible, rule-based keyword matching on assigner names, with overlaps between sectors (e.g., Open Source, Commercial Software, Web \& Content Management) resolved through a predefined prioritisation order. While this heuristic facilitates large-scale mapping, multi-sector entities and evolving business models may introduce classification uncertainties.

\begin{table}[htbp]
    \centering
    \rowcolors{3}{black!10}{white}
    \begin{tabular}{p{3.8cm}C{1.5cm}C{2cm}}
        \toprule
        Sector & All CVEs & Critical CVEs \\
        \midrule
        Cloud \& Hosting & 3700 & 263\\
        Commercial Software & 41583 & 4658\\
        Consulting \& Research & 82386 & 15744\\
        Consumer Electronics & 55 & 9\\
        Education \& Non-Profit & 33 & 4\\
        Finance \& Insurance & 66 & 2\\
        Hardware & 13474 & 1276\\
        Healthcare & 22 & 10\\
        Industrial \& IoT & 2493 & 308\\
        Open Source & 28699 & 2751\\
        Other & 51873 & 4055\\
        Platforms \& DevOps & 108 & 1\\
        Security Vendors & 7249 & 848\\
        Telecommunication \& Networking & 13620 & 1476\\
        Web \& Content Management & 95 & 25\\
        \bottomrule
    \end{tabular}
    \caption{Number of all and critical CVEs by sector.}
    \label{tab:sector_nos}
\end{table}
While sample sizes vary considerably across sectors (Table~\ref{tab:sector_nos}), the median-based analysis remains robust for identifying general remediation patterns, with smaller sectors (e.g., Healthcare, $n=22$) providing indicative trends rather than definitive benchmarks.

A further indication of this is provided by a comparison of performance across different sectors, which shows that open-source communities and cloud providers generally remediate faster than traditional industries, such as commercial software or hardware manufacturing. It is particularly evident in the healthcare, energy, and telecommunications sectors that there is a tendency for shorter patch cycles, which is likely attributable to the presence of more stringent legal and compliance requirements~\cite{bsi_itsig_wirksamkeit_2023, wef_cybersecurity_outlook_2025}.

\subsection{Case Study Insights}
The quantitative findings are reinforced by qualitative case studies of three landmark critical vulnerabilities (CVSS $\geq 9.0$), illustrating real-world manifestation of detection, disclosure, and remediation patterns observed across the 2009-2024 dataset.

The Heartbleed vulnerability (CVE-2014-0160), which was discovered in OpenSSL in April~2014, enabled attackers to read up to $64\,\text{KB}$ of server memory, with the potential to expose private keys, passwords and session data for millions of systems worldwide~\cite{nist_heartbleed_cve}. Despite the rapid provision of patches within two days, the delays between disclosure and remediation in the affected companies averaged more than six months. This was due to widespread dependency on the open-source library and lack of automated detection tools. This case study highlights the challenges associated with the ``last mile'' of patch deployment, emphasising the necessity for dependency scanning and automated updating within open-source ecosystems to mitigate prolonged exposure windows, which align with the quantitative medians (1,668 days for critical CVEs).

EternalBlue (CVE-2017-0144), which was exploited in the WannaCry ransomware of May~2017, targeted Windows SMBv1 protocol flaws, affecting unpatched Windows systems globally and causing damages in excess of \$4\,billion~\cite{nist_eternalblue_cve}. Microsoft released a patch in March 2017 (pre-disclosure), yet six months post-disclosure, $20\,\%$  of organisations remained vulnerable, thereby amplifying the subsequent ransomware outbreak. The incident demonstrates the protracted nature of remediation processes in commercial software sectors, extending beyond the established quantitative averages. This emphasises the necessity for regulatory mandates to enforce patch SLAs in critical infrastructures, such as healthcare and telecommunications.

The Log4Shell vulnerability (CVE-2021-44228), which was disclosed in December~2021 in Apache Log4j, enabled remote code execution via malicious logging inputs. This had a significant impact on more than 3 billion devices and triggered immediate zero-day exploits~\cite{nist_cve-2021-44228, bsi_cve-2021-44228, crowdstrike_log4shell}. Patches were issued within days; however, full remediation took weeks due to supply-chain propagation and configuration complexity, with global adoption lagging as documented in BSI warnings~\cite{bsi_cve-2021-44228}. This finding underscores the existence of persistent organisational delays, thereby signifying the necessity for continuous vulnerability management as opposed to periodic scans, as evidenced by the study's comprehensive patterns.

The collective analysis of these cases serves to reinforce the quantitative patterns identified, thereby illustrating how technical complexity, organisational inertia, and sectoral variations collectively drive the remediation gaps.

\section{Discussion And Evaluation}
The results of this study highlight both significant progress and persistent structural challenges in the management of critical software vulnerabilities. 

Over the period under scrutiny, the time lag between CVE reservation and public disclosure decreased significantly, reaching an average of approximately 33 days in 2024. This phenomenon points to an enhancement in the coordination and responsiveness of the global vulnerability management ecosystem. The expansion of the CNA programme and enhanced collaboration between security researchers, vendors, and vulnerability databases have likely contributed to this acceleration~\cite{cve_cna,cve_25_report}.The acceleration of disclosure processes has been demonstrated to increase transparency and enable organisations to initiate defensive measures earlier.

Despite these improvements, the findings demonstrate that faster disclosure does not necessarily lead to faster remediation. The analysis indicates that remediation timelines persistently exceed expectations and demonstrate considerable variability, with a median duration of 1,668 days for critical vulnerabilities. This discrepancy underscores a systemic "last-mile" problem in vulnerability management, where the primary bottleneck shifts from vulnerability discovery to the development and deployment of patches~\cite{alexiou_patch_management}.

The existence of this discrepancy can be attributed to a number of factors. In the contemporary context, software systems frequently employ complex dependency chains and interconnected components, a factor that has been shown to complicate the processes of patch development and testing. Moreover, organisational constraints, such as limited resources, operational risks associated with updates, and fragmented asset inventories, have the potential to delay patch deployment, particularly in large or legacy environments.

Sectoral comparisons provide further support for these observations. It is evident that open-source ecosystems and cloud-based platforms frequently demonstrate a higher level of responsiveness, largely attributable to the utilisation of automated development pipelines and continuous integration practices. Conversely, traditional commercial vendors characteristically implement more extended release cycles. It has been demonstrated that regulated sectors, including but not limited to healthcare and telecommunications, exhibit accelerated remediation, a phenomenon that is presumably precipitated by regulatory incentives.

Case studies, including those of Heartbleed, EternalBlue and Log4Shell, illustrate that even when patches are released promptly, organisations frequently require a considerable amount of time to identify affected systems and deploy updates across complex infrastructures. The findings emphasise that effective vulnerability mitigation is contingent not only on vendor response, but also on organisational preparedness and the implementation of mature patch management processes.

The findings indicate an enhancement in vulnerability management with regard to transparency and disclosure efficiency. Nevertheless, the extended remediation timelines underscore a persistent "last-mile" issue in the implementation of patch deployment. In order to address this challenge, it is necessary to implement not only faster vulnerability reporting but also improved automation, better asset visibility, and more mature patch management processes within organisations.

\section{Conclusion and Future Work}
The present study examined the processes of detection, disclosure, and remediation of critical vulnerabilities that had severity scores of nine or higher. Integration of data-driven analysis and qualitative case evaluations enabled identification of both structural improvements and persistent challenges in contemporary vulnerability management.\par

The findings indicate that global disclosure timelines have become considerably reduced, signifying a maturation of the ecosystem of coordinated vulnerability reporting and enhanced cross-organisational communication~\cite{cve_25_report, bsi_cvd_leitlinie}. Nevertheless, the remediation phase continues to demonstrate deficiencies, with notable heterogeneity in patch release times across software vendors and sectors. These delays, frequently attributable to resource constraints, legacy dependencies, and fragmented responsibilities, result in organisations remaining vulnerable for extended periods following the disclosure of vulnerabilities~\cite{alexiou_patch_management}.\par

This work contributes to extant research by quantifying the systemic inefficiencies that persist despite procedural advances. It is also important to note that effective vulnerability management is not just a technical problem, but also a governance challenge~\cite{bsi_itsig_wirksamkeit_2023}. In order to address this challenge, there is a need for synchronised policy, automation and human expertise.
Organisations that actively integrate regulatory frameworks, establish prioritised workflows and mandate security accountability are better positioned to reduce the window between discovery and mitigation.\par

From an applied perspective, this study underscores the necessity for organisations to accord priority to critical vulnerabilities (CVSS $\geq 9.0$) through the implementation of established SLAs (cf. Section~\ref{chap:related_work}) and sector-specific strategies (Subsection~\ref{subsec:organizational_and_sectorial_veriations}), while concomitantly addressing systemic remediation delays that have been identified across the period 2009 to 2024 (see Figures \ref{fig:cdf_patch_duration}, \ref{fig:patch_availability}, \ref{fig:patch_by_sector}), particularly in commercial software sectors that require a longer timeframe to remediate (cf. Figure \ref{fig:patch_by_sector}).

In future research, the exploration of machine learning–driven models, such as exploit prediction scoring systems, in the refinement of prioritisation in dynamic threat environments is recommended. Further investigation into the following areas would be of use in attempting to bridge the gap between awareness and action: automated remediation pipelines; CI/CD-integrated patching; and multi-source vulnerability aggregation~\cite{first_epss}. Furthermore, emerging paradigms, such as exposure management and zero-trust architectures offer promising frameworks for the unification of vulnerability management across hybrid infrastructures.\par

The study indicates that, while the speed of identifying and publishing critical vulnerabilities is improving, sustainable cybersecurity resilience depends on transitioning from reactive vulnerability management to proactive, continuous exposure governance.

\section*{Acknowledgement}
The present paper is founded upon Lena's Master Thesis, which was conducted at the Faculty of Computer Science and Electrical Engineering, Kiel University of Applied Sciences.

\balance
\printbibliography

@online{cve_faqs, 
    author = {{MITRE Corporation}},
    title = {Frequently Asked Questions (FAQs) - What is CVE?}, 
    publisher = {{MITRE Corporation}},
    url = {https://www.cve.org/ResourcesSupport/FAQs},
    urldate = {2026-03-14}
}

@online{nist_cvss_scoring, 
    author = {{National Institute of Standards and Technology (NIST)}},
    title = {Vulnerability Metrics}, 
    publisher = {{National Institute of Standards and Technology (NIST)}},
    url = {https://nvd.nist.gov/vuln-metrics/cvss#},
    urldate = {2026-03-14}
}

@online{ibm_cost_of_breach,
    author = {{Ponemon Institute}},
    title = {Cost of a Data Breach Report 2024},
    publisher = {{IBM}},
    pages = {14},
    year = {2024},
    url = {https://table.media/wp-content/uploads/2024/07/30132828/Cost-of-a-Data-Breach-Report-2024.pdf},
    urldate = {2026-03-14}
}

@online{bsi_lage_it-sicherheit_schwachstellen_wachstum,
    author = {{Bundesamt für Sicherheit in der Informationstechnik}},
    title = {The state of IT security in Germany in 2023 (Original title in German: {Die Lage der IT-Sicherheit in Deutschland 2023})},
    pages = {85-87},
    year = {2023},
    url = {https://www.bsi.bund.de/SharedDocs/Downloads/DE/BSI/Publikationen/Lageberichte/Lagebericht2023.pdf},
    urldate = {2026-03-14}
}

@online{nist_cve-2021-44228, 
    author = {{National Institute of Standards and Technology (NIST)}},
    title = {{CVE}-2021-44228 Detail}, 
    publisher = {{National Institute of Standards and Technology (NIST)}},
    url = {https://nvd.nist.gov/vuln/detail/CVE-2021-44228},
    urldate = {2026-03-14}
}

@online{bsi_cve-2021-44228,
    author = {{Bundesamt für Sicherheit in der Informationstechnik}},
    title = {Critical vulnerability published in log4j (CVE-2021-44228) (Original title in German: {Kritische Schwachstelle in log4j veröffentlicht (CVE-2021-44228)})}, 
    publisher = {{Bundesamt für Sicherheit in der Informationstechnik}},
    url = {https://www.bsi.bund.de/SharedDocs/Cybersicherheitswarnungen/DE/2021/2021-549032-10F2.pdf?__blob=publicationFile&v=5},
    urldate = {2026-03-14}
}

@online{crowdstrike_log4shell,
  author = {{CrowdStrike Intelligence Team}},
  title = {Log4j2 Vulnerability "Log4Shell" ({CVE}-2021-44228)},
  year = {2021},
  url = {https://www.crowdstrike.com/en-us/blog/log4j2-vulnerability-analysis-and-mitigation-recommendations/},
  urldate = {2026-03-14}
}

@online{first_cvss_v4_guide,
    author = {{Forum of Incident Response and Security Teams (FIRST)}},
    title = {Common Vulnerability Scoring System v4.0 – User Guide},
    publisher = {{Forum of Incident Response and Security Teams (FIRST)}},
    url = {https://www.first.org/cvss/v4.0/user-guide},
    urldate = {2026-03-14}
}

@online{bsi_patchmanagement_2021,
  author = {{Bundesamt für Sicherheit in der Informationstechnik (BSI)}},
  title = {{OPS}.1.1.3: Patch and change management (Original title in German: {{OPS}.1.1.3: Patch- und Änderungsmanagement})},
  year = {2021},
  url = {https://www.bsi.bund.de/SharedDocs/Downloads/DE/BSI/Grundschutz/IT-GS-Kompendium_Einzel_PDFs_2021/04_OPS_Betrieb/OPS_1_1_3_Patch_und_Aenderungsmanagement_Edition_2021.pdf?__blob=publicationFile&v=2},
  urldate = {2026-03-14}
}

@article{alexiou_patch_management,
    author = {Spiros Alexiou},
    title = {Practical Patch Management and Mitigation},
    journal = {{ISACA Journal}},
    volume = {2019},
    number = {3},
    year = {2019},
    pages = {1-6},
    url = {https://www.isaca.org/resources/isaca-journal/issues/2019/volume-3/practical-patch-management-and-mitigation},
    urldate = {2026-03-14}
}

@techreport{nist_patch_guide_2022,
  author = {Souppaya, Murugiah and Scarfone, Karen},
  title = {Guide to Enterprise Patch Management Planning: Preventive Maintenance for Technology},
  institution = {{National Institute of Standards and Technology (NIST)}},
  year = {2022},
  number = {{NIST SP} 800-40 Rev. 4},
  url = {https://nvlpubs.nist.gov/nistpubs/SpecialPublications/NIST.SP.800-40r4.pdf4},
  urldate = {2026-03-14}
}

@online{nurse_patch_challenges,
    author = {Jason R. C. Nurse},
    title = {To Patch or Not to Patch: Motivations, Challenges, and Implications for Cybersecurity},
    year = {2025},
    publisher = {{Institute of Cyber Security for Society and School of Computing, University of Kent}},
    url = {https://arxiv.org/pdf/2502.17703},
    urldate = {2026-03-14}
}

@article{przymus_secret_life_of_cves,
  author = {Przymus, Piotr and Fejzer, Mikołaj and Narębsk, Jakub and Stencel, Krzysztof},
  title = {The Secret Life of CVEs},
  journal = {arXiv preprint},
  year = {2025},
  url = {https://arxiv.org/pdf/2504.03863},
  urldate = {2026-03-14}
}

@article{przymus_transitive_lifecycle,
  author = {Przymus, Piotr and Fejzer, Mikołaj and Narębsk, Jakub and Rykaczewski, Krzysztof and Stencel, Krzysztof},
  title = {Out of Sight, Still at Risk: The Lifecycle of Transitive Vulnerabilities in Maven},
  journal = {arXiv preprint},
  year = {2025},
  url = {https://arxiv.org/pdf/2504.04803},
  urldate = {2026-03-14}
}

@online{bsi_itsig_wirksamkeit_2023,
  author    = {{Bundesamt für Sicherheit in der Informationstechnik (BSI)}},
  title     = {Study on the effectiveness of IT security laws among operators of critical infrastructures (Original title in German: {Untersuchung zur Wirksamkeit der IT-Sicherheitsgesetze unter Betreibern Kritischer Infrastrukturen})},
  year      = {2023},
  url       = {https://www.bsi.bund.de/SharedDocs/Downloads/DE/BSI/KRITIS/evaluierung-itsig2-ergebnisbericht.pdf?__blob=publicationFile&v=3},
  urldate = {2026-03-14}
}

@online{nist_cve_json2.0_data,
    author = {{National Institute of Standards and Technology (NIST)}},
    title = {{NVD} Data Feeds - {JSON} 2.0 Feeds},
    publisher = {National Institute of Standards and Technology (NIST))},
    url = {https://nvd.nist.gov/vuln/data-feeds},
    date = {2025-05-28},
    urldate = {2026-03-14}
}

@online{mitre_cve_json_data,
    author = {{MITRE Corporation}},
    title = {{CVE} List Downloads},
    publisher = {{MITRE Corporation}},
    url = {https://www.cve.org/Downloads},
    date = {2025-05-20},
    urldate = {2026-03-14}
}

@online{nist_heartbleed_cve,
  author = {{National Institute of Standards and Technology (NIST)}},
  title = {{CVE}-2014-0160 Detail},
  year = {2014},
  url = {https://nvd.nist.gov/vuln/detail/CVE-2014-0160},
  urldate = {2026-03-14}
}

@online{nist_eternalblue_cve,
  author = {{National Institute of Standards and Technology (NIST)}},
  title = {{CVE}-2017-0144 Detail},
  year = {2017},
  url = {https://nvd.nist.gov/vuln/detail/CVE-2017-0144},
  urldate = {2026-03-14}
}

@online{cve_cna, 
    author = {{MITRE Corporation}},
    title = {{CVE} Numbering Authority (CNA) Operational Rules}, 
    publisher = {MITRE Corporation},
    url = {https://www.cve.org/ResourcesSupport/AllResources/CNARules},
    urldate = {2026-03-14}
}

@online{cve_cna_list, 
    author = {{MITRE Corporation}},
    title = {List of Partners}, 
    publisher = {{MITRE Corporation}},
    note = {\url{https://www.cve.org/PartnerInformation/ListofPartners}},
    urldate = {2026-03-14}
}

@online{cve_glossar_record, 
    author = {{MITRE Corporation}},
    title = {Glossary - CVE Record}, 
    publisher = {{MITRE Corporation}},
    note = {\url{https://www.cve.org/ResourcesSupport/Glossary\#glossaryRecord}},
    urldate = {2026-03-14}
}

@online{cve_record_process, 
    author = {{MITRE Corporation}},
    title = {Process}, 
    publisher = {{MITRE Corporation}},
    note = {\url{https://www.cve.org/About/Process}},
    urldate = {2026-03-14}
}

@online{wef_cybersecurity_outlook_2025,
    author = {{World Economic Forum}},
    title = {Global Cybersecurity Outlook 2022},
    year = {2025},
    publisher = {World Economic Forum},
    url = {https://reports.weforum.org/docs/WEF_Global_Cybersecurity_Outlook_2025.pdf},
    urldate = {2026-03-14}
}

@online{cve_25_report,
  author = {{MITRE Corporation}},
  title = {{CVE}® 25 YEARS - 25th Anniversary Report October 2024},
  publisher = {{MITRE Corporation}},
  year = {2024},
  pages = {3-6},
  note = {\url{https://www.cve.org/Resources/Media/Cve25YearsAnniversaryReport.pdf}},
  urldate = {2026-03-14}
}

@online{bsi_cvd_leitlinie,
  author = {{Bundesamt für Sicherheit in der Informationstechnik (BSI)}},
  title = {BSI guideline on the Coordinated Vulnerability Disclosure (CVD) process (Original title in German: {Leitlinie des {BSI} zum Coordinated Vulnerability Disclosure ({CVD})-Prozess})},
  year = {2022},
  url = {https://www.bsi.bund.de/SharedDocs/Downloads/DE/BSI/CVD/CVD-Leitlinie.pdf?__blob=publicationFile&v=4},
  urldate = {2026-03-14}
}

@online{nist_nvd-process, 
    author = {{National Institute of Standards and Technology (NIST)}},
    title = {{CVE}s and the {NVD} Process}, 
    publisher = {{National Institute of Standards and Technology (NIST)}},
    url = {https://nvd.nist.gov/general/cve-process},
    year = {2024},
    urldate = {2026-03-14}
}

@online{first_epss,
    author = {{Forum of Incident Response and Security Teams (FIRST)}},
    title = {Exploit Prediction Scoring System (EPSS) - Frequently Asked Questions},
    publisher = {{Forum of Incident Response and Security Teams (FIRST)}},
    url = {https://www.first.org/epss/faq},
    urldate = {2026-03-14}
}

\end{document}